\documentclass[useAMS,usenatbib]{mn2e}
\usepackage{graphicx}
\usepackage{longtable}
\usepackage{auto-pst-pdf}

\title[A 20 GHz Bright Sample for $\delta > 72^{\circ}$-- II. Multi-frequency Follow-up]{A 20 GHz 
Bright Sample for $\delta > 72^{\circ}$-- II. Multi-frequency Follow-up}
\author[R. Ricci, S. Righini, R. Verma, and others]{R. Ricci$^{1}$\thanks{E-mail: ricci@ira.inaf.it}, 
S. Righini$^{1}$, R. Verma$^{2}$, I. Prandoni$^{1}$, E. Carretti$^{3}$, K.-H. Mack$^{1}$,
\newauthor
M. Massardi$^{1}$, P. Procopio$^{4,5}$, A. Zanichelli$^{1}$, L. Gregorini$^{1,8}$, F. Mantovani$^{1}$, 
\newauthor
M.P. Gawro\'nski$^{6}$, M.W. Peel$^{7}$ \\
$^{1}$ {\it INAF-IRA Bologna, Via Gobetti 101, I-40129, Bologna, Italy}\\
$^{2}$ {\it Department of Physics, Ramjas College, University of Delhi, Delhi, 110007, India}\\
$^{3}$ {\it CSIRO Astronomy and Space Science, PO Box 276, Parkes, NSW 2870, Australia}\\
$^{4}$ {\it INAF-IASF Bologna, Via Gobetti 101, I-40129, Bologna, Italy}\\
$^{5}$ {\it School of Physics, David Caro Building, Corner of Tin Alley and 
Swanston St, University of Melbourne, Parkville, VIC 3010, Australia }\\
$^{6}$ {\it Centre for Astronomy, Faculty of Physics, Astronomy and
Informatics, Nicolaus Copernicus University, Grudziadzka 5, 87-100 Toru\'n, Poland}\\
$^{7}$ {\it Jodrell Bank Centre for Astrophysics, The University of Manchester, Oxford Road, Manchester, M13 9PL, UK}\\
$^{8}$ {\it Department of Physics and Astronomy, University of Bologna, 
V.le Berti Pichat 6/2, 40127, Bologna, Italy }}

\begin{document}

\date{Accepted  . Received }

\pagerange{\pageref{firstpage}--\pageref{lastpage}} \pubyear{2013}

\maketitle

\label{firstpage}

\begin{abstract}
We present follow-up observations at 5, 8 and 30 GHz of the K-band Northern Wide Survey (KNoWS) 
20 GHz Bright Sample, performed with the 32-m Medicina Radio Telescope and the 32-m Toru\'n Radio Telescope.
The KNoWS sources were selected in the Northern Polar Cap ($\delta > 72^{\circ}$) and 
have a flux density limit S$_{\rm 20GHz} =$ 115 mJy. We include NVSS 1.4 GHz measurements 
to derive the  source radio spectra between 1.4 and 30 GHz. 
Based on optical identifications, 68 per cent of 
the sources are QSOs and 27 per cent are radio galaxies. 
A redshift measurement is available for 58 per cent of the sources. 
The radio spectral properties of the different source populations 
are found to be in agreement with those of other high-frequency selected samples.
\end{abstract}


\section{Introduction}
In recent years, several studies have been carried out on the radio spectral 
properties of extra-Galactic high-frequency selected radio sources. 

Sadler et al. (2006) presented the general properties of a sample of radio 
sources with flux densities above 100 mJy obtained as part of the pilot 
20~GHz survey (AT20G) undertaken with the Australia Telescope Compact Array
(Ricci et al. 2004). With the use of a radio ``two-colour diagram'', first
used in the radio band by Kesteven et al. (1977),
they characterized the spectral properties of the high-frequency radio sources 
and confirmed the diversity of the radio spectra and the difficulty of 
predicting high-frequency properties by extrapolating the results of 
surveys observed at lower frequencies (see also Tucci et al. 2008 and 
Massardi et al. 2008). Considering the two classical radio source populations 
(extended steep-spectrum and compact flat-spectrum sources), Sadler et al. (2006) 
found that roughly 87 per cent of the high-frequency selected sources are 
flat-spectrum and 13 per cent are steep-spectrum.  
Tucci et al. (2008) performed a  multifrequency spectral analysis of a sample of radio 
sources observed with the Very Small Array (VSA) at 33 GHz exploiting existing data at
different frequencies down to 1.4 GHz. One of the most interesting results 
they found is that the majority of 33~GHz-selected sources show flatter spectra 
going to higher frequency (about 50 per cent of the sample) or become inverted 
(about 19 per cent of the sample). This contradicts the standard models for 
high-frequency source spectra, which predict a spectral steepening due to 
the ageing of high-energy electrons. The result they found is completely 
consistent with the ``unified model'' for AGN: the steep-spectrum component, 
which arises from extended radio lobes, rapidly decreases with frequency, 
while the compact emission starts to dominate at frequency $\gg$ 1~GHz if the 
radio jet axis lies enough close to the line of sight. 
The difficulty of predicting high-frequency flux densities from the
 extrapolation of lower frequency data is testified by the significant number 
of sources found at high frequency that were not predicted. The percentage of these
``unexpected'' sources is observed to steadily increase with frequency: 10 per cent at 
15~GHz (Taylor et al. 2001), 18 per cent at 20~GHz (Sadler et al. 2006) and 
32 per cent at 33 GHz (Tucci et al. 2008).

The analysis of the radio spectral properties of 320 bright extragalactic radio 
sources with flux density $>$500~mJy at 20~GHz extracted from the AT20G was 
discussed in Massardi et al. (2008). In this sample, spectral steepening is a 
common feature that becomes more prominent at higher frequencies, even if flat 
and inverted spectrum objects still dominate the sample.

In a more recent paper, Prandoni et al (2010) studied the radio properties of 
a faint sample (S $>$0.6 mJy), selected at lower frequency (5~GHz) and 
associated with early-type galaxies, finding strong spectral similarities 
with the Massardi et al. (2008) sample and suggesting that 
faint radio galaxies are mostly counterparts of the brighter ones, with no 
significant influence introduced by different flux limits and frequency selection.

Massardi et al. (2011a) analysed the Full Sample Release 
(FSR, Murphy et al. 2010) of the AT20G survey, which is 91 per cent complete above 100~mJy and
79 per cent complete above 50~mJy, 
and found that the high-frequency bright source sample is dominated by 
flat-spectrum sources, while the fraction of steep-spectrum sources increases 
with decreasing flux density. These results have also been  
confirmed by Procopio et al. (2011), who observed 263 sources,
selected at 23~GHz in the WMAP maps, at 5, 8 and 22~GHz 
almost simultaneously to the observations performed between 30 and 857~GHz
by the European Space Agency Planck Satellite (Tauber et al. 2010). 
The optical properties of the AT20G full sample  presented in Mahony et al. 
(2011), who used SuperCOSMOS (Hambly et al. 2001) for optical identification 
and the 6DF Galaxy Survey (Jones et al. 2009) or the literature for redshifts, 
showed that the sources which accrete cold gas have steeper radio spectral 
indices with respect to the sources that accrete hot gas. This behaviour 
suggests that the radio emission in sources with cold-mode accretion is 
dominated by the jets. Interestingly, the fraction of flat- and 
steep-spectrum sources for the objects with no identification (blank field) 
is very different to that observed in the spectroscopic sample; in fact the 
percentage of blank fields increases dramatically at steep spectral indices. 
The authors suggested that this increase could indicate a population of 
high-z ultra-steep-spectrum sources.

In recent papers, Massardi et al. (2011b) and Bonavera et al. (2011) presented 
the Planck Australia Telescope Compact Array Co-eval Observations (PACO), 
which provided two complete samples of flux density-selected sources: a bright sample
($S_{20GHz} >$ 500 mJy) and a faint sample ($S_{20GHz} >$ 200 mJy) 
with extensive multi-frequency coverage (5--40~GHz). 
The spectral analysis showed no significant differences between the two samples.
The steepening of the sources at high frequencies is confirmed. The main 
difference is a larger fraction of steep-spectrum sources in the 
faint sample, mostly at the expense of peaked- and flat-spectrum sources. In a
subsequent paper, Bonaldi et al. (2013) presented a detailed analysis of a
complete sample of 69 sources spectrally-selected from the AT20G sample as
those having inverted or upturning spectra and observed in the framework
of the PACO project. Most of the source spectra (85 per cent) are smooth and well
described by a double power-law, while the remaining sources have complex 
spectra. The majority of the radio sources are likely blazars probably caught 
during a bright phase at 20~GHz.

Gawro\'nski et al. (2010) also investigated the dependence of spectral 
index distribution on flux density for a sample of 57 sources detected 
at 30~GHz with the OCRA prototype receiver at the 32-m Toru\'n radio telescope
above a flux density limit of 5 mJy. They found that the proportion of steep 
spectrum sources increases with decreasing flux density. They found no evidence 
for an unexpected population of sources above their completeness limit of 10 mJy 
whose spectra rise towards higher frequencies.   

Peel et al. (2011) measured the 30~GHz flux densities of 605 radio sources 
from the Combined Radio All-sky Target Eight-GHz Survey (CRATES) with the 
same instrumentation as Gawro\'nski et al. (2010). They studied the radio
spectra between 1.4 and 30~GHz and found that 75 per cent have steepened 
at $\alpha_{8.4}^{30}$ compared with $\alpha_{1.4}^{4.8}$, 6 per cent of the 
sources show a GHz-Peaked Spectrum shape with a peak between 4.8 and 8.4~GHz,
10 per cent of the sources are flat or rising and another 10 per cent are 
inverted.     

Kurinsky et al. (2013) reported the multi-frequency observations 
(4.8--43.3 GHz) of 89 sources with 37~GHz flux density $>$ 1 Jy carried out 
with the VLA/JVLA. The sample contains a much higher fraction (42 per cent) of 
flat-spectrum sources compared to 5--10 per cent of sources found in other samples 
selected at lower frequency. This sample extends the radio spectral energy 
distribution of the radio sources in the 5--857~GHz regime.   
 
In this paper we present the results of the multi-frequency follow-up observations of
the K-band Northern Wide Survey (KNoWS) Pilot project (Righini et al. 2012, 
hereafter Paper~I) in a comprehensive way. 
In Section~\ref{sez:bright} a brief summary of the pilot survey is provided; 
while in Section~\ref{sez:lowfol} the results of the 5/8~GHz
follow-up observations are presented. In Section~\ref{sez:ocra} we report on the 
30~GHz follow-up observations and data reduction. In Section~\ref{sez:spec} 
we show the radio spectra between 1.4 and 30~GHz and the optical identifications. 
In Section~\ref{sez:res} we present the results of the analysis of radio spectra, 
and we compare them with the results from other high-frequency selected samples. 
In Section~\ref{sez:sum} we summarize our conclusions.     
Throughout this paper we use the following cosmological parameters: $H_{\rm 0}=$71 km 
s$^{-1}$ Mpc$^{-1}$, $\Omega_{\rm m}=$0.27 and $\Omega_{\rm \lambda}=$0.73 (Larson et al. 2011).
        
\section[]{The KNoWS Bright source sample} \label{sez:bright}

During the years 2010--2011, the Medicina 32-m dish hosted the seven-feed 18--26.5~GHz receiver built 
for the Sardinia Radio Telescope, with the aim of performing its commissioning. This opportunity 
was exploited to carry out a pilot survey at 20~GHz over the sky region north of $\delta = 72.3^{\circ}$. 
This survey produced a catalogue of 73 confirmed sources down to a flux density limit of 115 mJy, 
which can be considered complete above 200 mJy. The 73 confirmed sources come from a selected
list of 151 candidates chosen from the brighter and more reliable targets for the 20~GHz 
follow-up. For this reason the sample analysed in this Paper is referred to as the KNoWS 
20~GHz Bright Sample. A more detailed description is provided in Paper~I. 

Precise 20~GHz flux densities and position measurements were derived for all sources through 20~GHz 
follow-up observations, carried out at the 32-m Medicina telescope, in the framework of the 
multi-frequency (5, 8, 20, 30~GHz) study of the sample. The Medicina 20~GHz follow-up observations 
are described in detail in Paper~I. Here we focus on the 5/8~GHz Medicina Observations and on 
the 30~GHz observations carried out at the Toru\'n telescope.   

\section[]{5/8~GHz follow-up observations} \label{sez:lowfol}

Follow-up observations of the KNoWS Bright Sample were performed with the 32-m Medicina 
Telescope on Dec 11$^{\rm th}$, 12$^{\rm th}$ and 13$^{\rm th}$ 2010 in the frequency ranges 
5--5.15~GHz and 8.18--8.33~GHz. The target list of 73 sources 
was scheduled to be observed at 5 and 8~GHz in blocks of
pointings based on RA range. A list of flux density calibrators (3C286, 3C295, 3C48,
3C123 and NGC7027) were observed twice a day interspersed with blocks of 
target sources. Both calibrators and target sources were observed in multiple 
RA and Dec cross-scans centred on the positions obtained from the KNoWS 20~GHz Bright Sample. 
The cross-scans were 6$\times$HPBW wide (where the Half Power Beam Width
is 7.5 arcmin at 5~GHz and 4.8 arcmin at 8.3~GHz) at the scanning speed of 
3$^{\circ}$/min  and 2$^{\circ}$/min  at 5 and 8~GHz respectively. 
For comparison, the angular resolution at 20~GHz is 1.7 arcmin. Data 
were recorded into FITS files. They were then flagged
in order to discard scans clearly affected by Radio Frequency 
Interference and/or severe atmospheric disturbances. The data were  
flux-calibrated according to the absolute flux density scale by Ott et al. (1994). 
The RA and Dec cross-scans were then stacked for each target source
in order to improve the S/N ratio of the detections by decreasing the rms 
noise. The stacked cross-scan was Gaussian-fitted in
order to obtain flux density and positional offsets in RA and Dec. 
The positional offsets were then used to correct the flux density for the 
signal attenuation caused by the primary beam profile. A thorough description 
of the aforementioned calibration and fitting procedures (performed with the data 
reduction package OSCaR, Procopio et al. 2011), together with the error analysis 
for the flux density measurements are provided in Paper~I. 

Of the 73 sources, 60 (82 per cent) and 57 (78 per cent) were detected at 5 and 8~GHz 
respectively. Non-detections at 5/8~GHz were mainly due to the limited sensitivity 
of the receivers, which were not properly cooled down at the time of observations.
The $5\sigma$ flux density limit was 63 mJy/beam at 5 GHz and 85 mJy/beam at 8 GHz.
The 5/8~GHz flux densities of the sources are listed in Table~\ref{tab:flux}.    
                     
\begin{figure} 
\begin{center}
 \includegraphics[height=8cm]{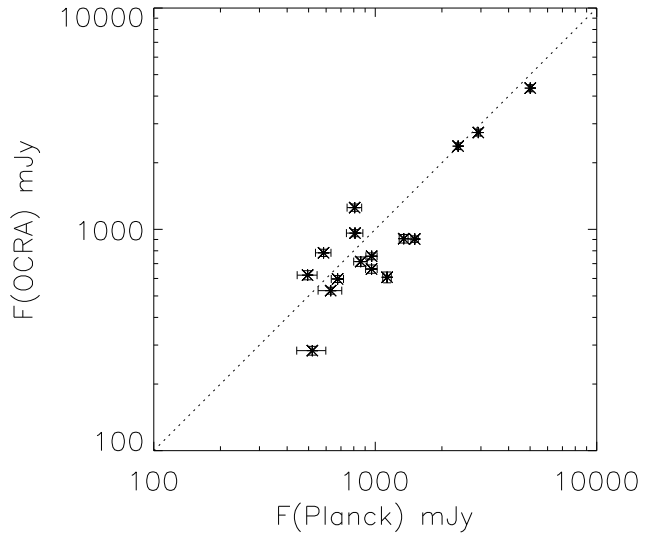} 
\end{center}
 \caption{Scatter plot between the 30~GHz Planck ERCSC flux densities and 
30~GHz OCRA flux densities with error bars.} \label{fig:ocra}
\end{figure}
     
\section[]{30~GHz follow-up observations} \label{sez:ocra}

The 30~GHz follow-up observations were carried out with the OCRA-p receiver at 
the Toru\'n 32-m telescope (Poland). OCRA-p is the prototype of the OCRA-F 
(One Centimeter Radio Array-Faraday) multi-feed 30~GHz receiver developed in the 
framework of the OCRA program (Browne et al. 2000).  The nominal sensitivity of 
the receiver considering an overall system temperature of 40~K and a bandwidth of 
6~GHz, is 6 mJy s$^{1/2}$. The FWHM of the antenna beam is 72 arcsec at 30~GHz. 
The 30 GHz observations were carried out on September 9$^{th}$, November
15$^{th}$ and December 3$^{rd}$ 2010. 
The targets were observed using cross-scans or on/off measurements depending 
on the source flux density (Lowe et al. 2007; Gawro\'nski et al. 2010).
The planetary nebula NGC~7027 was used as primary flux density calibrator. 
NGC~7027 was observed 6--10 times for each observing session. Hafez et al. 
(2008) report a flux density of 5.39$\pm$0.04~Jy at 33~GHz at an epoch of 
2003.0 with a secular decrease of $-0.17\pm0.03$ per cent per year (Ott et al.
 1994). By extrapolating the flux density to 2010.0 a value of 5.32$\pm$0.04~Jy
was obtained, which was then scaled to 30 GHz using the NGC~7027 spectral index
of $-0.119$. The 30~GHz flux density of NGC~7027 thus obtained is 
5.38$\pm$0.04 Jy. Secondary flux density calibration was performed using the
signal generated from a noise diode after each source observation. The flux
density of NGC~7027 was measured at different elevations in order to obtain 
elevation-dependent gain corrections. The correction for atmospheric absorption
was obtained by calculating the opacity from the system temperature measurements
at zenith and at 30$^{\circ}$ of Elevation.

The data reduction was performed using a custom software package described in Peel et al. (2011) 
and in greater detail in Peel (2009). 
In total 295 measurements were obtained on the full list of 73 sources. Data affected by 
poor weather conditions were discarded, as well as cross-scans in which the peak amplitude 
measured by the OCRA-p beams was offset by more than 20 per cent. 

52 targets were detected at 30~GHz corresponding to 71 per cent of the total. The 30~GHz 
flux densities of the detected sources are listed in Table~\ref{tab:flux}.

Fig.~\ref{fig:ocra} shows the scatter plot between the 30~GHz Planck 
Early Release Compact Source Catalogue (ERCSC; Planck Collaboration 2011) flux 
densities and the 30~GHz OCRA flux densities of the 16 targets present in both data sets. 
Planck and OCRA flux densities are in reasonable agreement especially at the bright end, 
indicating that our flux densities are not affected by significant systematic effects.         
Source variability might account for a scatter that is larger than the flux density 
error bars for the fainter objects. It might also be useful to compare in terms of 
variability the Planck-OCRA scatter plot with the one in Figure 6 of Gawro\'nski et al. 
(2010) where 42 OCRA-p 30~GHz flux densities are matched with those from the 
Very Small Array source subtractor at 33~GHz. 10 out of the 42 sources show variations
greater than 2$\sigma$ (combined error).

\section[]{Radio Spectra and Optical Identifications} \label{sez:spec}

The radio spectra of the KNoWS sources are shown in Fig.~\ref{fig:SED}. 
Integrated flux densities at 1.4~GHz were added from the NRAO-VLA Sky Survey 
(NVSS, Condon et al. 1998) to extend the frequency baseline. The NVSS flux densities 
were extracted from the NVSS online catalogue from a cross-match with the KNoWS 20~GHz 
follow-up positions within a search radius of 50~arcsec. In order to assess the 
reliability of these cross-matches, all the 1.4~GHz NVSS images were retrieved from the 
NVSS online postage-stamp server and visually inspected. 

\begin{figure*}
  \includegraphics[width=15cm]{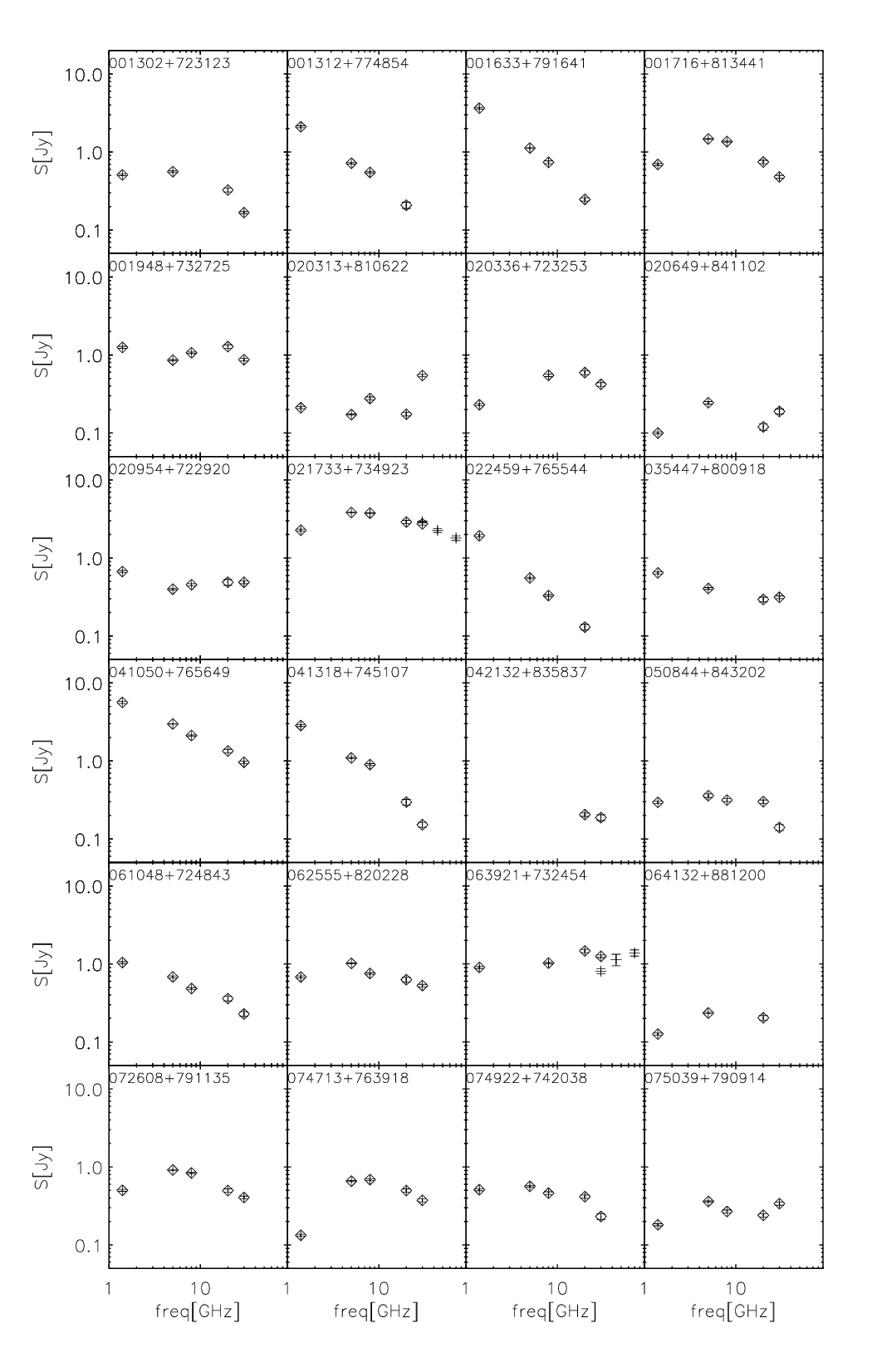}
 \caption{Radio spectra of the KNoWS sample between 1.4~GHz and 30~GHz obtained using the flux densities
in Table~\ref{tab:flux} (diamonds) together with the 30, 44 and 70~GHz Planck ERCSC flux densities (crosses) where available.} \label{fig:SED}
\end{figure*}
\addtocounter{figure}{-1}
\begin{figure*}
  \includegraphics[width=15cm]{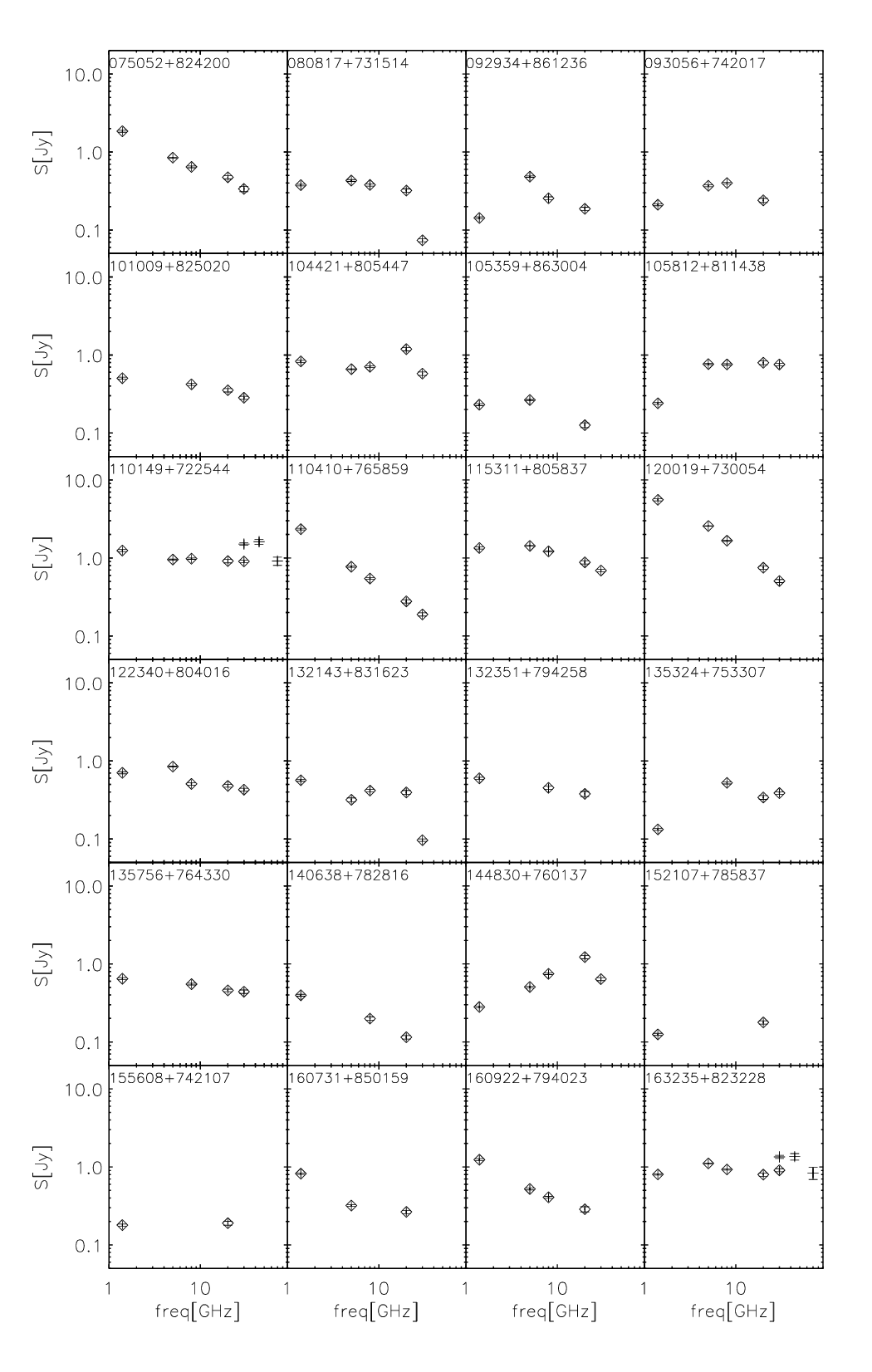}
  \caption{Continued.}
\end{figure*}
\addtocounter{figure}{-1}
\begin{figure*}
  \includegraphics[width=15cm]{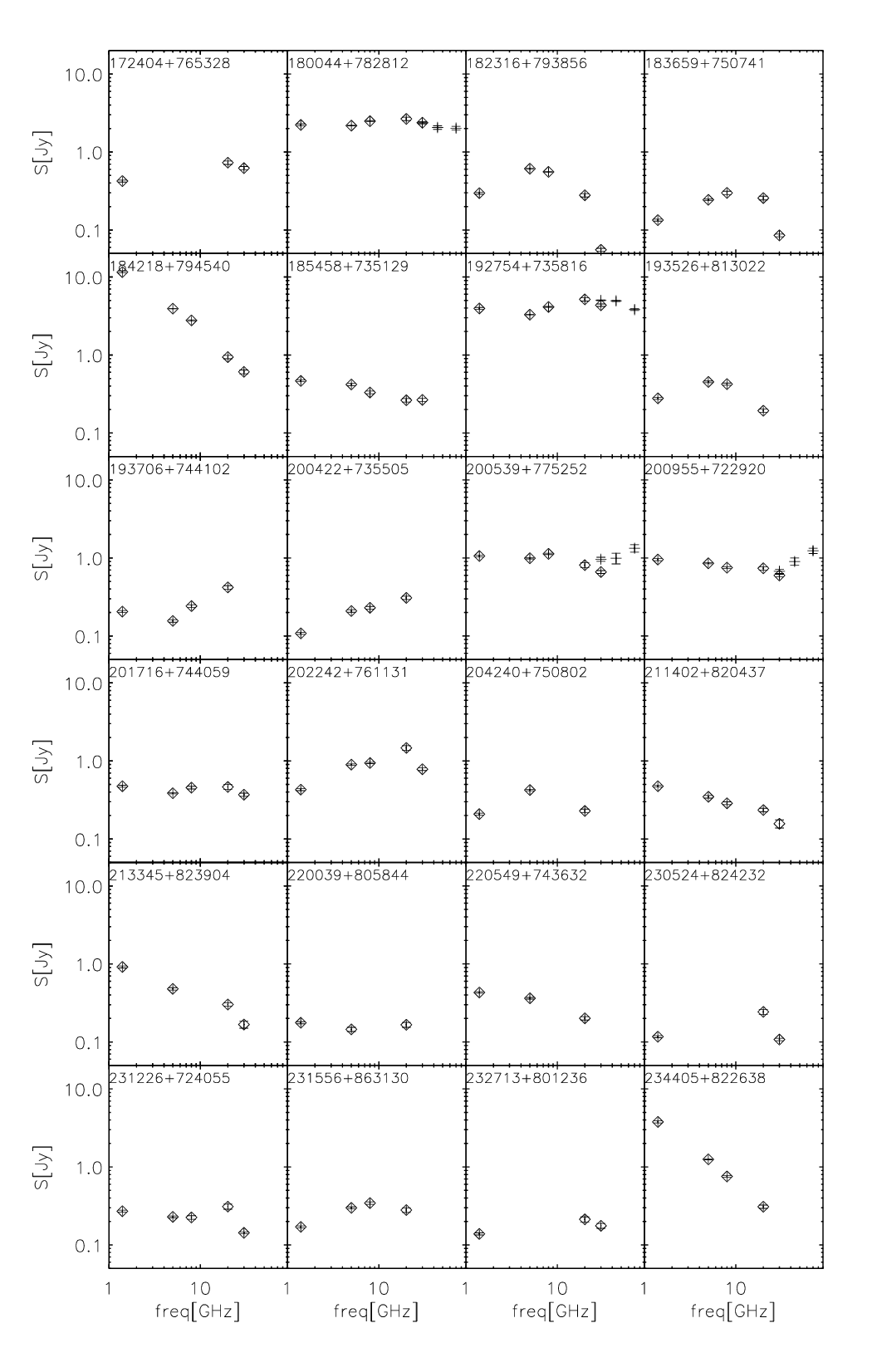}
  \caption{Continued.}
\end{figure*}

All 73 sources in our sample appear to have NVSS counterparts, with three
sources (KNoWS~080817+731514, KNoWS~144830+760137, KNoWS~184218+794540) showing 
two NVSS components and three sources (KNoWS~160731+850159, KNoWS~211402+820437, 
KNoWS~220549+743632) showing three NVSS components in the NVSS images. For these six 
sources the NVSS components were summed to provide the integrated 1.4~GHz flux densities 
reported in Table~\ref{tab:flux}.

When available, the 30~GHz, 44~GHz and 70~GHz Planck ERCSC (Planck Collaboration 2011) 
flux densities were also added to the radio spectra. However such measurements 
are not included in Table~\ref{tab:flux} due to the sparsity of this information. 

We note that flux density measurements used to 
derive the radio spectra are not coeval. In particular the high-frequency ones ($\ge$ 5~GHz) 
were taken within a time range of 1.5 years: between August 2009 and June 2010 for the Planck flux 
density measurements; in November 2010 for the OCRA-p 30~GHz KNoWS follow-up; in December 
2010 for the 5/8~GHz follow-up and in April 2011 for the KNoWS 20~GHz follow-up measurements.
The 1.4 GHz NVSS observations were done between September 1993 and October 1996 with additional patching 
observations carried out during the fourth quarter of 1997.
This means that our radio spectra might be affected by variability.   
Massardi et al. (2010) found a 5--6 per cent median variability on a 6 months
timescale in the range 5--18 GHz, and up to a $\sim9$ per cent variability over 9
months, so we do not expect variability to significantly affect the spectral analysis discussed in
the following sections. The long time lag between the 1.4 GHz observations and the higher 
frequency ones (17 years) is not troublesome because at lower frequencies most of the emission 
comes from radio lobes which are more stable than radio cores.

In this paper we are interested only to the statistical behaviour of spectral indices in 
the sample. In this case positive and negative variations caused by short-term variability tend to cancel out. 
However, it must be born in mind, when using our data to interpret individual source spectra, that some 
sources may experience large variability also in relatively short periods. Some long-term 
flux density monitoring projects carried out at high frequencies at OVRO, Metsahovi and UMRAO 
(Richards et al. 2011, Ciaramella et al. 2004, Pyatunina et al. 2005, Hovatta et al. 2007) 
show plenty of such examples.

The NASA Extragalactic Database (NED; {\it http://ned.ipac.caltech.edu/}) was then 
used to identify the radio source optical counterparts, to find redshifts and to obtain optical types 
within a search radius of 1 arcmin to the 20~GHz KNoWS follow-up positions. 
In cases of multiple counterparts we chose the one which matched the NVSS flux density 
and position information. 

We also extracted the information on the optical class, b$_{\rm J}$ and r$_{\rm F}$ magnitude 
from the SuperCosmos Sky Surveys database (SSS; {\it http://www-wfau.roe.ac.uk/sss/index.html}). 
We again allowed for a 1-arcmin wide search box and in cases of multiple counterparts 
we chose the entry with the minimum distance from the follow-up 20~GHz radio position. 
The optical and redshift information of the KNoWS sample is summarized in Table~\ref{tab:optid}.  

According to NED we have 36 QSOs (49 per cent), 17 galaxies (23 per cent) and one Planetary Nebula, 
while 19 objects (36 per cent) have no optical types. In SuperCosmos, 18 objects (25 per cent) show galaxy 
appearance (class 1), 51 (70 per cent) a stellar appearance (class 2), one (the Planetary Nebula) is 
noisy (class=4) and 3 objects (4 per cent) have no SuperCosmos identification. For the 19 sources 
that have no optical type in NED we use the SuperCosmos classification,when available. 
Putting together the NED and SuperCosmos classifications, sources in the KNoWS Bright Sample 
can be divided as follows: 20 (27 per cent) are Galaxies or with galaxy appearance and 50 (68 per cent) 
are QSOs or with stellar appearance, in agreement with the results of previous works 
(see e.g. Sadler et al. 2006). Two have no optical classification and one is a Planetary Nebula. 

42 out of the 72 (58 per cent) extragalactic objects (we exclude here the Planetary Nebula) in the 
KNoWS Bright Sample have a redshift measurement.
30 of them are QSOs and 12 are radiogalaxies according to the combined NED/SuperCosmos optical 
classification. For 34 objects (24 QSOs and 10 radiogalaxies) we computed the  
radio and optical luminosities using the scheme described in Mahony et al. (2011). 
To compute luminosities across a wide range of redshifts it is essential to shift the observed 
frequencies into the rest frame using a $k-$correction. As QSOs and galaxies have very different 
SEDs in the optical regime each optical class needs to be $k-$corrected in a different manner. 
The $k-$correction formula by de Propris et al. (2004), making use of the b$_{\rm J}-$r$_{\rm F}$ colour, 
was applied to the radiogalaxies. For QSOs a power law was instead used with a spectral index $\beta = 0.5$.  

The redshift and luminosity statistics for the QSOs and radiogalaxies in the KNoWS sample are 
summarized in Table~\ref{tab:lumz}. As expected, QSOs are found on average at higher redshifts than 
radiogalaxies and are intrinsically brighter both in the radio and the optical band.      

\onecolumn

\begin{longtable}{lrrrrrrrrrr}

\caption{Col.(1) source name; 
cols.(2-3) the 30~GHz OCRA flux densities with errors; 
cols.(4-5) the 20~GHz follow-up flux densities with errors; 
cols.(6-7) the 8~GHz follow-up flux densities with errors; 
cols.(8-9) the 5~GHz follow-up flux densities with errors; 
cols.(10-11) the 1.4~GHz NVSS flux densities with errors.} \label{tab:flux} \\
\hline\hline \\
(1)  & (2) & (3)              & (4) & (5)              & (6)& (7)              & (8)& (9)             & (10) & (11) \\
Name & S30 & $\sigma_{\rm S30}$ & S20 & $\sigma_{\rm S20}$ & S8 & $\sigma_{\rm S8}$ & S5 & $\sigma_{\rm S5}$ & S1.4 & $\sigma_{\rm S1.4}$  \\
     &(mJy)&  (mJy)           & (mJy)& (mJy)             & (mJy)& (mJy)            &(mJy)& (mJy)            & (mJy)& (mJy) \\ 
\hline
\endfirsthead
\caption{continued.}\\
\hline\hline
(1)  & (2) & (3)              & (4) & (5)              & (6)& (7)              & (8)& (9)             & (10) & (11) \\
Name & S30 & $\sigma_{\rm S30}$ & S20 & $\sigma_{\rm S20}$ & S8 & $\sigma_{\rm S8}$ & S5 & $\sigma_{\rm S5}$ & S1.4 & $\sigma_{\rm S1.4}$  \\
     &(mJy)&  (mJy)           & (mJy)& (mJy)             & (mJy)& (mJy)            &(mJy)& (mJy)            & (mJy)& (mJy) \\ 
\hline
\endhead
\hline
\endfoot

KNoWS 001302$+$723123 &  167    &   5    &   325 &    18 & --  & --  &   559   &    14 &   509.1&  11.2  \\   
KNoWS 001312$+$774854 &  --   &  --  &   208 &    19 &   548 &    11 &   718   &    16 &  2101.3&  63.0  \\   
KNoWS 001633$+$791641 &  --   &  --  &   247 &    12 &   738 &    31 &  1123   &    14 &  3651.0&  88.6  \\   
KNoWS 001716$+$813441 &  480    &  20    &   745 &    38 &  1362 &    19 &  1471   &    16 &   692.8&  20.8  \\   
KNoWS 001948$+$732725 &  870    &  31    &  1283 &    64 &  1068 &    25 &   858   &    14 &  1251.8&  37.6  \\   
KNoWS 020313$+$810622 &  546    &  17    &   174 &     9 &   277 &    12 &   172   &     3 &   211.0&   6.3  \\   
KNoWS 020336$+$723253 &  420    &  22    &   594 &    34 &   549 &    21 & --    & --  &   229.9&   6.9  \\   
KNoWS 020649$+$841102 &  189    &  14    &   120 &    12 & --  & --  &   244   &     9 &   100.0&   3.0  \\   
KNoWS 020954$+$722920 &  488    &  21    &   488 &    42 &   454 &    14 &   398   &     8 &   669.9&  20.1  \\   
KNoWS 021733$+$734923 & 2739    &  96    &  2894 &   145 &  3760 &    32 &  3829   &    22 &  2271.6&  68.1  \\   
KNoWS 022459$+$765544 &  --   &  --  &   130 &    11 &   330 &     9 &   554   &    12 &  1927.2&  57.8  \\   
KNoWS 035447$+$800918 &  313    &  13    &   294 &    21 & --  & --  &   406   &    11 &   643.7&  19.3  \\   
KNoWS 041050$+$765649 &  962    &  36    &  1342 &    67 &  2122 &    38 &  2984   &    19 &  5620.1& 168.6  \\     
KNoWS 041318$+$745107 &  152    &  11    &   297 &    25 &   900 &    22 &  1094   &    18 &  2847.6&  80.0  \\   
KNoWS 042132$+$835837 &  188    &  16    &   205 &    10 & --  & --  & --    & --  &     5.2&   0.6  \\   
KNoWS 050844$+$843202 &  140    &  12    &   301 &    15 &   315 &    13 &   359   &    19 &   294.8&   8.9  \\   
KNoWS 061048$+$724843 &  228    &  16    &   360 &    25 &   484 &    12 &   683   &    18 &  1041.6&  31.3  \\   
KNoWS 062555$+$820228 &  528    &  19    &   628 &    45 &   755 &    17 &  1016   &    14 &   681.0&  20.4  \\   
KNoWS 063921$+$732454 & 1254    &  48    &  1467 &    73 &  1026 &    22 & --    & --  &   903.6&  27.1  \\   
KNoWS 064132$+$881200 &  --   &  --  &   204 &    14 & --  & --  &   235   &     5 &   126.2&   3.8  \\   
KNoWS 072608$+$791135 &  405    &  17    &   498 &    25 &   836 &    18 &   914   &    14 &   501.0&  15.0  \\   
KNoWS 074713$+$763918 &  373    &  17    &   498 &    25 &   688 &    17 &   657   &    11 &   133.1&   4.0  \\   
KNoWS 074922$+$742038 &  231    &  19    &   416 &    21 &   461 &    19 &   563   &    11 &   510.3&  15.3  \\    
KNoWS 075039$+$790914 &  339    &  16    &   240 &    12 &   268 &    12 &   361   &     6 &   181.5&   5.5  \\   
KNoWS 075052$+$824200 &  336    &  26    &   475 &    24 &   644 &    16 &   846   &    11 &  1845.1&  55.4  \\   
KNoWS 080817$+$731514 &   74    &   3    &   320 &    16 &   378 &    13 &   431   &    12 &   376.8&   9.3  \\   
KNoWS 092934$+$861236 &  --   &  --  &   187 &     9 &   255 &    11 &   483   &    10 &   142.7&   4.3  \\   
KNoWS 093056$+$742017 &  --   &  --  &   240 &    15 &   400 &    11 &   369   &    10 &   211.0&   6.3  \\   
KNoWS 101009$+$825020 &  283    &  13    &   353 &    18 &   420 &    13 & --    & --  &   503.7&  15.1  \\   
KNoWS 104421$+$805447 &  577    &  30    &  1188 &    56 &   706 &    17 &   658   &     7 &   828.3&  24.9  \\  
KNoWS 105359$+$863004 &  --   &  --  &   126 &     9 & --  & --  &   265   &     4 &   230.6&   6.9  \\   
KNoWS 105812$+$811438 &  756    &  27    &   795 &    44 &   756 &    17 &   764   &    13 &   240.3&   7.2  \\     
KNoWS 110149$+$722544 &  904    &  31    &   912 &    46 &   980 &    20 &   954   &    12 &  1245.6&  37.4  \\   
KNoWS 110410$+$765859 &  189    &   9    &   278 &    14 &   545 &    13 &   773   &    15 &  2340.6&  58.8  \\   
KNoWS 115311$+$805837 &  689    &  28    &   880 &    44 &  1215 &    23 &  1429   &    14 &  1343.4&  40.3  \\   
KNoWS 120019$+$730054 &  507    &  23    &   750 &    38 &  1666 &    30 &  2572   &    14 &  5564.7& 166.9  \\   
KNoWS 122340$+$804016 &  429    &  18    &   479 &    24 &   508 &    19 &   847   &    11 &   705.1&  21.2  \\   
KNoWS 132143$+$831623 &   96    &   3    &   395 &    28 &   417 &    17 &   318   &    18 &   565.4&  17.0  \\   
KNoWS 132351$+$794258 &  --   &  --  &   378 &    19 &   453 &    22 & --    & --  &   599.4&  18.0  \\   
KNoWS 135324$+$753307 &  387    &  16    &   341 &    17 &   524 &    17 & --    & --  &   132.6&   4.0  \\   
KNoWS 135756$+$764330 &  441    &  23    &   458 &    23 &   551 &    14 & --    & --  &   647.2&  19.4  \\   
KNoWS 140638$+$782816 &  --   &  --  &   115 &     6 &   199 &    10 & --    & --  &   396.8&  11.1  \\   
KNoWS 144830$+$760137 &  639    &  28    &  1225 &    61 &   747 &    17 &   507   &    13 &   280.8&   6.2  \\   
KNoWS 152107$+$785837 &  --   &  --  &   178 &     9 & --  & --  & --    & --  &   124.9&   3.8  \\   
KNoWS 155608$+$742107 &  --   &  --  &   190 &    11 & --  & --  & --    & --  &   179.9&   5.4  \\   
KNoWS 160731$+$850159 &  --   &  --  &   265 &    13 & --  & --  &   320   &     8 &   819.8&  18.0  \\   
KNoWS 160922$+$794023 &  --   &  --  &   287 &    22 &   408 &    12 &   522   &    15 &  1238.9&  37.2  \\   
KNoWS 163235$+$823228 &  908    &  40    &   798 &    40 &   928 &    15 &  1109   &    13 &   801.8&  14.8  \\   
KNoWS 172404$+$765328 &  621    &  29    &   730 &    37 & --  & --  & --    & --  &   424.0&  12.7  \\   
KNoWS 180044$+$782812 & 2376    &  81    &  2660 &   133 &  2482 &    44 &  2180   &    15 &  2223.5&  66.7  \\   
KNoWS 182316$+$793856 &   56    &   2    &   279 &    14 &   556 &     9 &   611   &    11 &   296.9&   8.4  \\   
KNoWS 183659$+$750741 &   85    &   3    &   257 &    14 &   299 &     6 &   244   &     5 &   133.9&   4.0  \\   
KNoWS 184218$+$794540 &  607    &  34    &   941 &    47 &  2773 &    48 &  3927   &    18 & 11555.9& 280.6  \\   
KNoWS 185458$+$735129 &  266    &  13    &   263 &    14 &   331 &    13 &   419   &    13 &   465.1&  11.7  \\   
KNoWS 192754$+$735816 & 4347    & 160    &  5165 &   258 &  4133 &    51 &  3289   &    23 &  3950.9& 118.5  \\  
KNoWS 193526$+$813022 &  --   &  --  &   193 &    10 &   423 &    13 &   452   &    12 &   278.1&   8.4  \\   
KNoWS 193706$+$744102 &  --   &  --  &   420 &    21 &   242 &    10 &   156   &     6 &   205.5&   6.2  \\   
KNoWS 200422$+$735505 &  --   &  --  &   308 &    15 &   230 &     9 &   209   &     8 &   108.0&   3.1  \\   
KNoWS 200539$+$775252 &  662    &  28    &   809 &    60 &  1128 &    23 &   993   &    20 &  1060.7&  29.8  \\   
KNoWS 200955$+$722920 &  597    &  22    &   737 &    37 &   752 &    16 &   855   &    12 &   953.6&  28.6  \\   
KNoWS 201716$+$744059 &  369    &  18    &   464 &    34 &   455 &    18 &   386   &     7 &   473.7&  14.2  \\   
KNoWS 202242$+$761131 &  784    &  29    &  1471 &   101 &   942 &    19 &   898   &    17 &   428.9&  12.9  \\   
KNoWS 204240$+$750802 &  --   &  --  &   229 &    12 & --  & --  &   424   &     9 &   208.3&   5.8  \\   
KNoWS 211402$+$820437 &  157    &  19    &   235 &    12 &   287 &    12 &   347   &    13 &   474.9&  11.2  \\   
KNoWS 213345$+$823904 &  167    &  17    &   302 &    15 & --  & --  &   479   &    17 &   915.1&  27.5  \\   
KNoWS 220039$+$805844 &  --   &  --  &   166 &    12 & --  & --  &   145   &     9 &   176.3&   5.3  \\   
KNoWS 220549$+$743632 &  --   &  --  &   201 &    10 & --  & --  &   363   &     9 &   430.5&   8.9  \\   
KNoWS 230524$+$824232 &  107    &   4    &   243 &    17 & --  & --  & --    & --  &  116.7 &  3.5   \\   
KNoWS 231226$+$724055 &  143    &   3    &   310 &    23 &   225 &    12 &   229   &     6 &   270.6&   8.1  \\   
KNoWS 231556$+$863130 &  --   &  --  &   280 &    19 &   344 &    13 &   299   &     8 &   170.5&   4.7  \\   
KNoWS 232713$+$801236 &  176    &  13    &   213 &    16 & --  & --  & --    & --  &   138.4&   4.2  \\   
KNoWS 234405$+$822638 &  --   &  --  &   309 &    15 &   755 &    16 &  1251   &    16 &  3777.4& 113.3  \\   
KNoWS 235626$+$815255 &  715    &  35    &   664 &    34 &   652 &    17 &   531   &     9 &   520.9&  15.6  \\   
\hline
\end{longtable}



\begin{longtable}{llllllcl}
\caption{Optical identification: 
col(1) source name;
col(2) radio spectral type as defined in Fig.~\ref{fig:3flux}; 
col(3) morphological type from NED;
col(4) redshift from NED;
col(5) b$_j$ magnitude from SuperCosmos;
col(6) r$_f$ magnitude from SuperCosmos;
col(7) Class from SuperCosmos (1: galaxy, 2: stellar);
col(8) other source name from NED.} \label{tab:optid} \\
\hline\hline \\
(1)  & (2)        & (3)  & (4) & (5)  & (6)  & (7)   & (8)       \\
Name & Radio Spectrum & NED ID & z   & b$_{j}$ & r$_{f}$ & Class & alt. name \\
\hline
\endfirsthead
\caption{continued.}\\
\hline\hline
(1)  & (2)        & (3)  & (4) & (5)  & (6)  & (7)   & (8)       \\
Name & Radio Spectrum & NED ID & z   & b$_{j}$ & r$_{f}$ & Class & alt. name \\
\hline
\endhead
\hline
\endfoot

KNoWS 001302$+$723123 &       flat & PN  &   ...     &    12.744 &   12.741 &   4 &   NGC0040                 \\              
KNoWS 001312$+$774854 &      steep & GAL &   0.326   &    22.351 &   --   &   1 &   NVSSJ001311$+$774846      \\            	
KNoWS 001633$+$791641 &      steep & GAL &   0.8404  &    21.707 &   19.731 &   2 &   3C 006.1                \\            	
KNoWS 001716$+$813441 &     peaked & QSO &   3.366   &    19.013 &   17.432 &   2 &   0014$+$813                \\            	
KNoWS 001948$+$732725 &       flat & QSO &   1.718   &    19.284 &   18.364 &   2 &   0016$+$731                \\            	
KNoWS 020313$+$810622 &      steep & NO  &   ...     &    19.952 &   18.621 &   2 &   NVSSJ020307$+$810612      \\            	
KNoWS 020336$+$723253 &   inverted & QSO &   ...     &    20.185 &   19.528 &   1 &   NVSSJ020333$+$723254      \\            	
KNoWS 020649$+$841102 &     peaked & NO  &   ...     &    21.837 &   --   &   1 &   NVSSJ020713$+$841119      \\            	
KNoWS 020954$+$722920 &       flat & GAL &   0.895   &    19.557 &   18.085 &   2 &   S5 0205$+$72              \\            	
KNoWS 021733$+$734923 &       flat & QSO &   2.367   &    17.090 &   15.298 &   2 &   0212$+$735                \\            	
KNoWS 022459$+$765544 &      steep & GAL &   ...     &    20.534 &   19.086 &   2 &   4C$+$76.01                \\            	
KNoWS 035447$+$800918 &       flat & NO  &   ...     &    19.831 &   --   &   2 &   0345$+$7958               \\            	
KNoWS 041050$+$765649 &      steep & GAL &   0.5985  &    21.678 &   --   &   2 &   4C$+$76.03                \\            	
KNoWS 041318$+$745107 &      steep & GAL &   0.373   &    20.811 &   18.782 &   1 &   4C$+$74.08                \\            	
KNoWS 042132$+$835837 &       -- & NO  &   ...     &    21.580 &   19.171 &   2 &   NVSSJ042140$+$835842      \\            	
KNoWS 050844$+$843202 &       flat & QSO &   1.340   &    20.134 &   19.116 &   2 &   0454$+$844                \\            	
KNoWS 061048$+$724843 &       flat & QSO &   3.53    &    22.337 &   --   &   2 &   4C$+$72.10                \\            	
KNoWS 062555$+$820228 &       flat & QSO &   0.710   &    21.368 &   18.946 &   1 &   0615$+$820                \\            	
KNoWS 063921$+$732454 &       flat & QSO &   1.85    &    18.592 &   18.172 &   2 &   NVSSJ063921$+$732458      \\            	
KNoWS 064132$+$881200 &       flat & NO  &   ...     &    20.291 &   --   &   1 &   NVSSJ064206$+$881154      \\            	
KNoWS 072608$+$791135 &     peaked & NO  &   ...     &    21.908 &   --   &   2 &   S5 0718$+$79              \\            	
KNoWS 074713$+$763918 &     peaked & NO  &   ...     &    12.224 &   10.830 &   2 &   S5 0740$+$76              \\            	
KNoWS 074922$+$742038 &       flat & QSO &   1.629   &    19.282 &   18.990 &   2 &   S5 0743$+$74              \\            	
KNoWS 075039$+$790914 &     peaked & NO  &   ...     &    21.917 &   --   &   2 &   NVSSJ075043$+$790917      \\            	
KNoWS 075052$+$824200 &      steep & QSO &   1.991   &    --   &   --   &   - &   0740$+$8249               \\            	
KNoWS 080817$+$731514 &       flat & QSO &   0.496   &    21.208 &   19.286 &   2 &   S5 0802$+$73              \\            	
KNoWS 092934$+$861236 &     peaked & QSO &   ...     &    21.563 &   --   &   1 &   S5 0916$+$86              \\            	
KNoWS 093056$+$742017 &     peaked & NO  &   ...     &    21.792 &   --   &   2 &   S5 0925$+$74              \\            	
KNoWS 101009$+$825020 &       flat & GAL &   0.322   &    20.792 &   19.672 &   1 &   S5 1003$+$83              \\            	
KNoWS 104421$+$805447 &     upturn & QSO &   1.26    &    18.685 &   17.978 &   2 &   1039$+$811                \\            	
KNoWS 105359$+$863004 &     peaked & GAL &   ...     &    21.253 &   --   &   1 &   NVSSJ105421$+$862936      \\            	
KNoWS 105812$+$811438 &   inverted & GAL &   0.706   &    19.470 &   19.036 &   2 &   S5 1053$+$81              \\            	
KNoWS 110149$+$722544 &       flat & QSO &   1.46    &    17.612 &   17.122 &   2 &   1058$+$726                \\            	
KNoWS 110410$+$765859 &      steep & QSO &   0.3115  &    15.330 &   14.978 &   2 &   PG1100$+$772              \\            	
KNoWS 115311$+$805837 &       flat & QSO &   1.25    &    18.787 &   18.173 &   2 &   1150$+$812                \\            	
KNoWS 120019$+$730054 &      steep & GAL &   0.97    &    20.133 &   18.369 &   1 &   3C268.1                 \\            	
KNoWS 122340$+$804016 &       flat & QSO &   ...     &    21.975 &   --   &   2 &   S5 1221$+$80              \\            	
KNoWS 132143$+$831623 &       flat & NO  &   1.024   &    20.508 &   18.712 &   2 &   NVSSJ132145$+$831614      \\            	
KNoWS 132351$+$794258 &       flat & QSO &   1.970   &    21.501 &   --   &   1 &   S5 1323$+$79              \\            	
KNoWS 135324$+$753307 &     peaked & QSO &   1.619   &    18.386 &   17.995 &   2 &   NVSSJ135323$+$753258      \\            	
KNoWS 135756$+$764330 &       flat & QSO &   ...     &    22.535 &   --   &   2 &   S5 1357$+$76              \\            	
KNoWS 140638$+$782816 &      steep & GAL &   ...     &    20.338 &   18.655 &   2 &   NVSS J140636$+$782810     \\            	
KNoWS 144830$+$760137 &   inverted & QSO &   0.899   &    21.996 &   --   &   1 &   S5 1448$+$76              \\            	
KNoWS 152107$+$785837 &       -- & NO  &   ...     &    18.996 &   18.382 &   2 &   NVSS J152102$+$785830     \\            	
KNoWS 155608$+$742107 &       -- & QSO &   1.667   &    20.435 &   19.641 &   2 &   NVSS J155603$+$742057     \\            	
KNoWS 160731$+$850159 &      steep & GAL &   0.183   &    18.072 &   16.660 &   1 &   S5 1616$+$85              \\            	
KNoWS 160922$+$794023 &      steep & NO  &   ...     &    --   &   --   &   - &   4C$+$79.15                \\            	
KNoWS 163235$+$823228 &       flat & GAL &   0.02471 &    14.531 &    7.384 &   1 &   NGC6251                 \\            	
KNoWS 172404$+$765328 &       -- & QSO &   0.680   &    18.901 &   18.428 &   2 &   1725$+$7655               \\            	
KNoWS 180044$+$782812 &       flat & QSO &   0.68    &    18.186 &   16.632 &   2 &   1803$+$784                \\            	
KNoWS 182316$+$793856 &     peaked & QSO &   0.224   &    19.269 &   17.415 &   1 &   S5 1826$+$79              \\            	
KNoWS 183659$+$750741 &       flat & NO  &   ...     &    19.798 &   20.297 &   2 &   1838$+$7504               \\            	
KNoWS 184218$+$794540 &      steep & GAL &   0.0561  &    12.582 &   11.015 &   2 &   3C390.3                 \\            	
KNoWS 185458$+$735129 &       flat & QSO &   0.461   &    16.521 &   16.411 &   2 &   S5 1856$+$73              \\            	
KNoWS 192754$+$735816 &       flat & QSO &   0.3021  &     7.445 &   16.557 &   2 &   4C $+$73.18               \\               	
KNoWS 193526$+$813022 &     peaked & GAL &   ...     &    19.060 &   17.912 &   2 &   S5 1939$+$81              \\            	
KNoWS 193706$+$744102 &     upturn & NO  &   ...     &    22.221 &   --   &   1 &   NVSS J193702$+$744054     \\            	
KNoWS 200422$+$735505 &   inverted & NO  &   ...     &    --   &   --   &   - &   NVSS J200417$+$735505     \\            	
KNoWS 200539$+$775252 &       flat & QSO &   0.342   &    14.039 &   13.372 &   2 &   2007$+$777                \\            	
KNoWS 200955$+$722920 &       flat & QSO &   ...     &    20.623 &   19.475 &   2 &   4C$+$72.28                \\            	
KNoWS 201716$+$744059 &       flat & QSO &   2.187   &    18.464 &   18.165 &   2 &   4C$+$74.25                \\            	
KNoWS 202242$+$761131 &   inverted & QSO &   ...     &    20.265 &   19.087 &   2 &   S5 2023$+$76              \\            	
KNoWS 204240$+$750802 &     peaked & QSO &   0.104   &    17.132 &   16.152 &   1 &   4C$+$74.26                \\            	
KNoWS 211402$+$820437 &       flat & GAL &   0.084   &    17.137 &   15.832 &   1 &   S5 2116$+$81              \\            	
KNoWS 213345$+$823904 &      steep & QSO &   2.357   &    21.414 &   19.450 &   2 &   S5 2136$+$82              \\            	
KNoWS 220039$+$805844 &       flat & NO  &   ...     &    21.534 &   20.664 &   2 &   2201$+$8043               \\            	
KNoWS 220549$+$743632 &       flat & GAL &   ...     &    21.532 &   20.585 &   2 &   S5 2205$+$74              \\            	
KNoWS 230524$+$824232 &       -- & QSO &   ...     &    18.390 &   17.336 &   2 &   S5 2304$+$82              \\            	
KNoWS 231226$+$724055 &       flat & NO  &   ...     &    18.827 &   17.855 &   2 &   NVSSJ231219$+$724127      \\            	
KNoWS 231556$+$863130 &       flat & NO  &   ...     &    20.992 &   18.721 &   2 &   NVSSJ231549$+$863143      \\            	
KNoWS 232713$+$801236 &       -- & NO  &   ...     &    22.170 &   20.618 &   2 &   NVSSJ232706$+$801259      \\            	
KNoWS 234405$+$822638 &      steep & QSO &   0.735   &    21.769 &   20.233 &   2 &   2342$+$821                \\            	
KNoWS 235626$+$815255 &       flat & QSO &   1.344   &    21.155 &   18.584 &   2 &   S5 2353$+$81              \\            	    
\end{longtable}
 
\twocolumn

\begin{table}
 \centering
  \caption{Redshift and luminosity statistics for the KNoWS sources 
with redshift information.} \label{tab:lumz}
  \begin{tabular}{lccc}
  \hline
                           &   QSO            &      Gal       &     Total       \\
  \hline
  Number                   &   30/50 (60\%)   &  12/20 (60\%)  &  42/72 (58\%)   \\        
   $<z>$                   &   1.33           &   0.45         &    --         \\
  \hline
  Number                   &  24/50 (48\%)    &  10/20 (50\%)   &   34/72 (48\%)  \\  
  $<\log L_{20} (W/Hz)>$    &  27.13           &  26.14         &    --         \\  
  $<M_{\rm bJ}>$             &  $-24.95$        & $-22.10$       &    --         \\
  \hline
\end{tabular}
\end{table}

\section[]{Spectral index analysis} \label{sez:res}

  

\begin{table*} 
 \centering
  \caption{Statistics of the spectral indices between 1.4 and 30~GHz.} \label{tab:alp}
  \begin{tabular}{lcccc}
  \hline
  $\alpha$                  & $\alpha_{1.4}^{5}$ & $\alpha_{5}^{8}$ & $\alpha_{8}^{20}$ & $\alpha_{20}^{30}$   \\
  \hline
mean($\alpha$)$\pm\sigma$   & $-0.01\pm 0.06$  & $-0.18\pm 0.08$ & $-0.27\pm0.07$   & $-0.80\pm 0.16$    \\
median($\alpha$)            & $-0.08$          & $-0.19$         & $-0.22$          & $-0.56$             \\
Source Number               & 67               & 50              & 50               & 52                  \\
  
  \hline
\end{tabular}
\end{table*}

The flux density measurements described in the previous sections and reported
in Table~\ref{tab:flux} were used to compute spectral indices $\alpha$ defined as $S \propto \nu^{\alpha}$. 
Four spectral indices were calculated in four different frequency ranges: 1.4--5~GHz $\alpha_{1.4}^{5}$; 
5--8~GHz $\alpha_{5}^{8}$; 8--20~GHz $\alpha_{8}^{20}$ and 20--30~GHz $\alpha_{20}^{30}$. 
Six sources (all classified as QSOs) 
have no flux density measurements at both 5~GHz and 8~GHz; for these sources we cannot compute 
the spectral indices, so they are excluded from the following statistical analysis,
which is therefore based on 67 radio sources. 
Table~\ref{tab:alp} shows how the radio spectra change while moving from 
lower (1.4~GHz) to higher (30~GHz) frequencies. A clear steepening trend is present
both in the mean and the median spectral index; this effect is in agreement 
with what was found by Massardi et al. (2011a) for the AT20G FSR 
(Murphy et al. 2010) and further investigated by Chhetri et al. (2012).
The median spectral indices in PACO Bright Sample are
$\alpha_{5}^{10}$ = 0.04 (while the faint sample was $-0.04$, Bonavera et al. 2011); 
$\alpha_{10}^{20} = -0.19$ and $\alpha_{20}^{30} = -0.45$. 
 
We used the radio spectra to compare the spectral properties of different 
source populations (QSOs and radiogalaxies) with the ones found in other 
high-frequency selected radio catalogues. 
The most useful tool in this respect is the colour-colour plot already used 
in the radio band by Sadler et al. (2006) to evaluate the spectral properties
of the AT20G Pilot Survey sample (Ricci et al. 2004). 








Sources occupy a place on the
colour-colour plot according to their spectral behaviour. Synchrotron
emitting objects typically maintain a steep spectrum over all the whole
frequency range and share the third quadrant with compact steep spectrum
objects that are associated with young objects. The synchrotron 
self-absorption sources occupy the central region of the plot, where the flat spectrum
is typically described as the superposition of compact self-absorbing
Doppler boosted components. Gigahertz Peaked Spectra (O'Dea 1998) with emission peaking
above 10 GHz occupy the fourth quadrant and, in case of the younger
objects, might show inverted spectrum (i.e. lying in the first quadrant).
Flaring blazars are distinguishable from the GPS only because of their
variability, faster for the most energetic and close to the nuclear region
flares, which also tend to peak at the higher frequency, occupying also
the first and second quadrants in the colour-colour plot.
Chhetri et al. (2012) confirmed that $\alpha = -0.5$ is a physically meaningful
threshold to distinguish compact, self-absorbed sources (roughly speaking
``flat spectrum sources'') from structurally complex extended objects
(``steep''), and that this selection is more effective the lower the
frequency range is in which it is applied.

\begin{figure} 
\begin{center}
 \includegraphics[height=8cm]{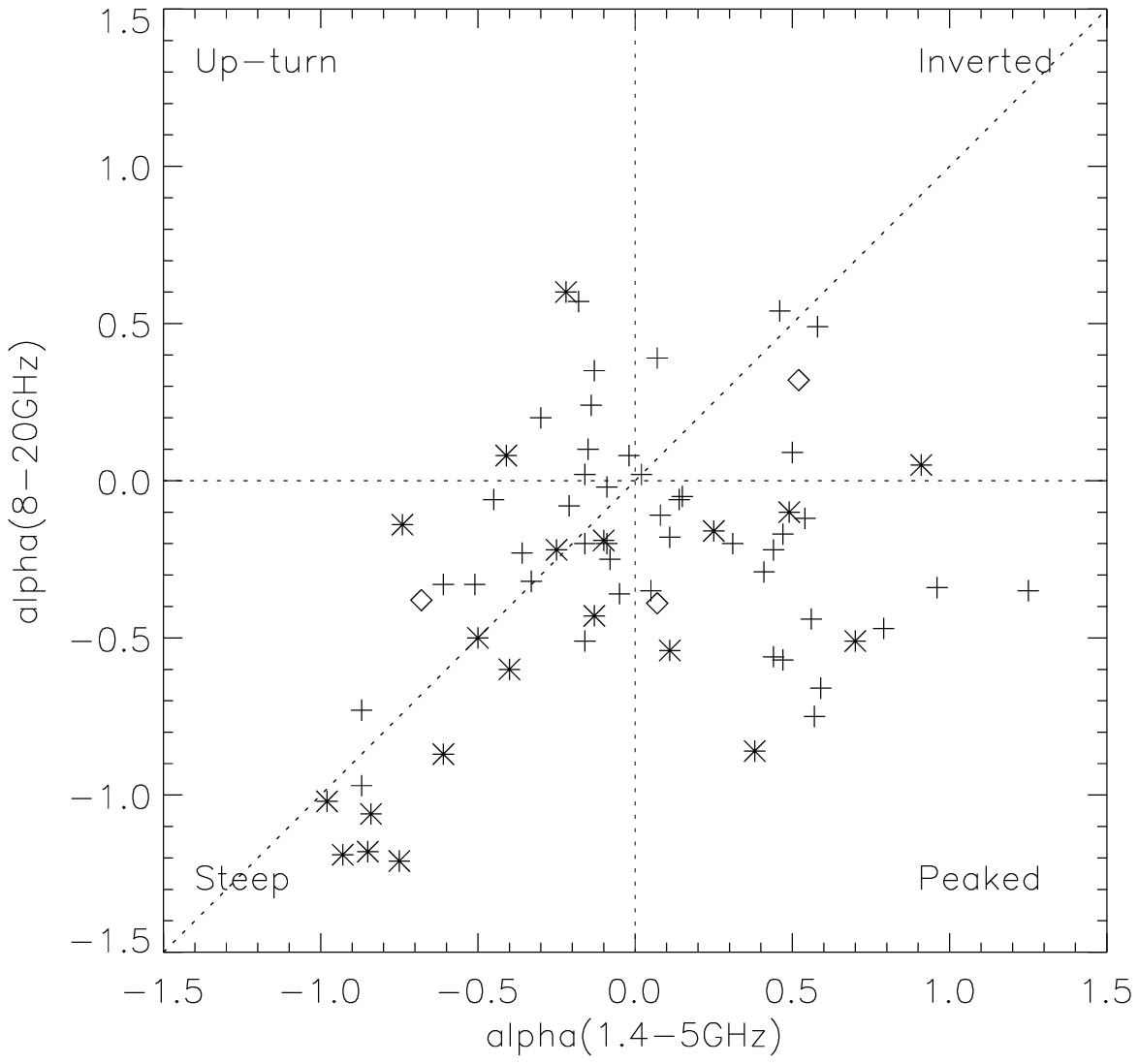} 
\end{center}
 \caption{Colour-colour plot: $\alpha_{8}^{20}$ vs $\alpha_{1.4}^{5}$. QSOs are indicated by crosses, 
radiogalaxies by asterisks and sources without classification and the Planetary Nebula by diamonds. 
The 1$-$1 line represents a {\it single} power-law spectrum. The spectral index 
error bars have the size of the symbols.} \label{fig:ccopt}
\end{figure}

We combined the $\alpha_{1.4}^{5}$ and $\alpha_{8}^{20}$ values for the 67 sources
of our sample with flux densities measured at 1.4, 5, 8 and 20~GHz to
create the colour-colour scatter plot 
shown in Fig.~\ref{fig:ccopt}. Radio sources of different optical types (according 
to our NED/SuperCosmos classification) are coded with different symbols.
It is clear from Fig.~\ref{fig:ccopt} that the radio galaxies mostly appear 
in the two lower quadrants, thus showing a declining radio spectrum going toward 
higher frequencies, while QSOs present a large scatter around the origin of the 
axes thus indicating that flat-spectrum radio quasars are more easily singled out 
in these high-frequency selected surveys. 

The statistics of the source populations appearing in the four quadrants of the 
colour-colour plot are detailed in Table~\ref{tab:souquad}, where they are compared 
to the statistics obtained from the AT20G FSR (Murphy et al. 2010), 
which covers the entire southern sky with a similar flux density limit. 
We notice that the statistics of the two samples is in good agreement especially
when taking into account the large statistical errors of our relatively small sample.

The values computed for our sample are then compared with other radio samples selected 
at different frequencies and with different flux density cut-offs 
(see Table~\ref{tab:soustats}): the sample of 
Massardi et al. (2008) selected at 20~GHz (S$_{\rm lim} = 500$ mJy), 
the PACO Bright (Massardi et al. 2011b) and Faint (Bonavera et al. 2011) samples
selected at 20~GHz (S$_{\rm lim} = 500$ mJy and 200 mJy, respectively),
the sample of Sadler et al. (2006) selected at 18~GHz (S$_{\rm lim} = 100$ mJy), 
the sample of Tucci et al. (2008) selected at 33~GHz (S$_{\rm lim} = 20$ mJy) and the  
AT20G FSR (Murphy et al. 2010) with S$_{\rm lim} = 50$ mJy.


\begin{table*} 
 \centering
  \caption{Statistics of the KNoWS source populations in the four 
quadrants of $\alpha_{8}^{20}$ vs $\alpha_{1.4}^{5}$ colour-colour plot. Source fractions 
for the AT20G Full Sample Release are reported for comparison in square brackets.} \label{tab:souquad}
  \begin{tabular}{lcc}

  \hline
  spectral type &  QSOs  &  radiogalaxies \\     
  \hline
  steep         & 14 (52$\pm$15\%) [38\%]   &  12 (44$\pm$11\%) [32\%] \\
  peaked        & 18 (75$\pm$17\%) [52\%]   &  5  (21$\pm$8\%) [21\%]  \\
  inverted      & 5 (71$\pm$29\%) [43\%]    &  1  (14$\pm$14\%) [30\%] \\
  up-turn       & 7 (77$\pm$33\%) [45\%]    &  2  (22$\pm$11\%) [29\%]  \\
  \hline
\end{tabular}
\end{table*}

Even if a clear increasing trend in the fraction 
of steep-spectrum sources with decreasing flux density in Table~\ref{tab:soustats} 
does not seem statistically significant, it is true that steep-spectrum sources may be 
over-represented in the two lowest flux limit samples indicating that a change might be present in the 
dominant source population at flux densities of $<100$ mJy. It is also worth noticing 
that most of the up-turned spectra in Fig.~\ref{fig:ccopt} have shifted to the inverted 
source quadrant. This effect might be caused by different spectral shapes in the low 
1.4--5~GHz frequency domain with respect to the 5--8~GHz domain (e.g. due to the extended 
features such as radio lobes dominating the low frequency observations).

The KNoWS sample also appears to be in substantial agreement with the AT20G FSR 
(Murphy et al. 2010) statistics of the different radio spectral types 
when the latter sample is cut at a similar flux density limit (100 mJy, see column 4). 
When the entire AT20G FSR catalog is analysed down to the limiting flux density of 50 mJy 
the percentage of steep-spectrum sources becomes significantly higher than the KNoWS 
percentage, this at the expense of inverted-spectrum and peaked-spectrum 
populations. On the other hand we notice that the KNoWS and the AT20G FSR statistics differ 
from those of Sadler et al. (2006) for up-turned, peaked and inverted-spectrum populations, with
Sadler et al. (2006) showing a higher percentage of sources 
even if all samples have the same flux density threshold.        

\begin{table*} 
 \centering
  \caption{Percentage of source populations (upturn, inverted, steep and 
peaked) with Poissonian error bars in different radio samples. 
The samples compared are KNoWS: this paper; M: Massardi et al. (2008);  
PACO Bright: Massardi et al. (2011b); PACO Faint: Bonavera et al. (2011); Sad: Sadler et al. (2006);  
AT20G FSR: Murphy et al. (2010) for two flux density limits (100 and 50 mJy); 
Tuc: Tucci et al. (2008). For KNoWS the spectral indices $\alpha_{1.4}^{5}$ 
and $\alpha_{8}^{20}$ were used to discriminate between source populations.} \label{tab:soustats}
  \begin{tabular}{lcccccccc}
  \hline
  Sample              & KNoWS     & M         &  PACO Bright    &  PACO Faint         & Sad        &  AT20G FSR  & AT20G FSR  &  Tuc          \\   
  S$_{\rm lim}$ (mJy)   & 115       & 500       &    500          &       200           & 100        &   100       &   50       &  20           \\     
  \hline                                                                                                                      
  upturn  [\%]        & 13$\pm$4  & 10$\pm$1  &  15$\pm$ 3      &       22 $\pm$ 4    & 22$\pm$5   &  13$\pm$1   &  13$\pm$1  & 29$\pm$6      \\
  inverted [\%]       & 10$\pm$4  & 28$\pm$5  &  $<$1           &        $<$1         & 20$\pm$4   &  13$\pm$1   &   9$\pm$1  & 7$\pm$3       \\
  steep   [\%]        & 37$\pm$7  & 30$\pm$5  &  45 $\pm$ 6     &       42 $\pm$ 5    & 34$\pm$6   &  43$\pm$1   &  50$\pm$1  & 53$\pm$7      \\
  peaked  [\%]        & 39$\pm$8  & 32$\pm$5  &  31 $\pm$ 5     &       29 $\pm$ 4    & 24$\pm$5   &  31$\pm$1   &  28$\pm$1  & 11$\pm$3      \\
  \hline                                                                      
  Total source no.    & 67        & 218       &  174            &       143           & 101        &   2001      &  3698      &  102          \\
  \hline
\end{tabular}
\end{table*}

\begin{table*} 
 \centering
  \caption{Percentage comparison between the KNoWS sample, the AT20G FSR  
and the PACO Bright and Faint samples for different spectral source populations and flux 
density ranges. In the KNoWS column the number of objects is also given in round brackets. 
For KNoWS and AT20G FSR, $\alpha_{5}^{8}$ and $\alpha_{8}^{20}$ were used to discriminate 
between source populations. Another source class (``flat'') was introduced for this table, 
which was not present in the previous two tables.} \label{tab:at20g}
  \begin{tabular}{lccccccc}
  \hline
  Sample             &  KNoWS               &   FSR         &    PACO Bright    &   PACO Faint     &  FSR          & FSR          \\
  S$_{\rm lim}$ (mJy)  &  115                 &   500         &         500       &      200         &  100          & 50           \\ 
  \hline                                                                                                        
  Inverted           & (3)  6$\pm$2       &  11$\pm$2   &     7  $\pm$ 2    &      4 $\pm$ 1   & 9$\pm$1     & 6$\pm$0.5   \\
  Peaked             & (3)  6$\pm$2       &  8$\pm$2    &     5  $\pm$ 1    &      2 $\pm$ 1   & 7$\pm$1     & 6$\pm$0.5  \\
  Upturn             & (0)  $<$4          &  $<$1       &     $<$1          &        $<$1      & 2$\pm$0.3   & 3$\pm$0.3   \\
  Steep              & (20) 40$\pm$6      &  12$\pm$2   &     14 $\pm$ 3    &     14 $\pm$ 3   & 27$\pm$1    & 33$\pm$1    \\
  Flat               & (24) 48$\pm$7      &  68$\pm$5   &     74 $\pm$ 6    &     80 $\pm$ 7   & 54$\pm$2    & 53$\pm$1    \\
  \hline                                                                                                        
  Total source no.   & 50                   & 254           &     174           &     143          & 2019          & 3332           \\ 
  \hline
\end{tabular}
\end{table*}

To better characterize the flat-spectrum source population, Massardi et al. (2011a) divided 
the sources of the AT20G FSR with flux densities measured at 5, 8 and 20 GHz into five classes 
based on the place they occupy in a $\alpha_{8}^{20}$ vs $\alpha_{5}^{8}$ colour-colour 
plot, according to the following scheme:
   
\begin{itemize}

\item Flat: $-0.5 < \alpha_{5}^{8} < 0.5$ and $-0.5 < \alpha_{8}^{20} < 0.5$

\item Inverted: $\alpha_{5}^{8} > 0$ and $\alpha_{8}^{20} > 0$ minus flat region 

\item Peaked: $\alpha_{5}^{8} > 0$ and $\alpha_{8}^{20} < 0$ minus flat region
    
\item Upturn: $\alpha_{5}^{8} < 0$ and $\alpha_{8}^{20} > 0$ minus flat region

\item Steep: $\alpha_{5}^{8} < 0$ and $\alpha_{8}^{20} < 0$ minus flat region

\end{itemize} 
 
\begin{figure} 
\begin{center}
 \includegraphics[height=8cm]{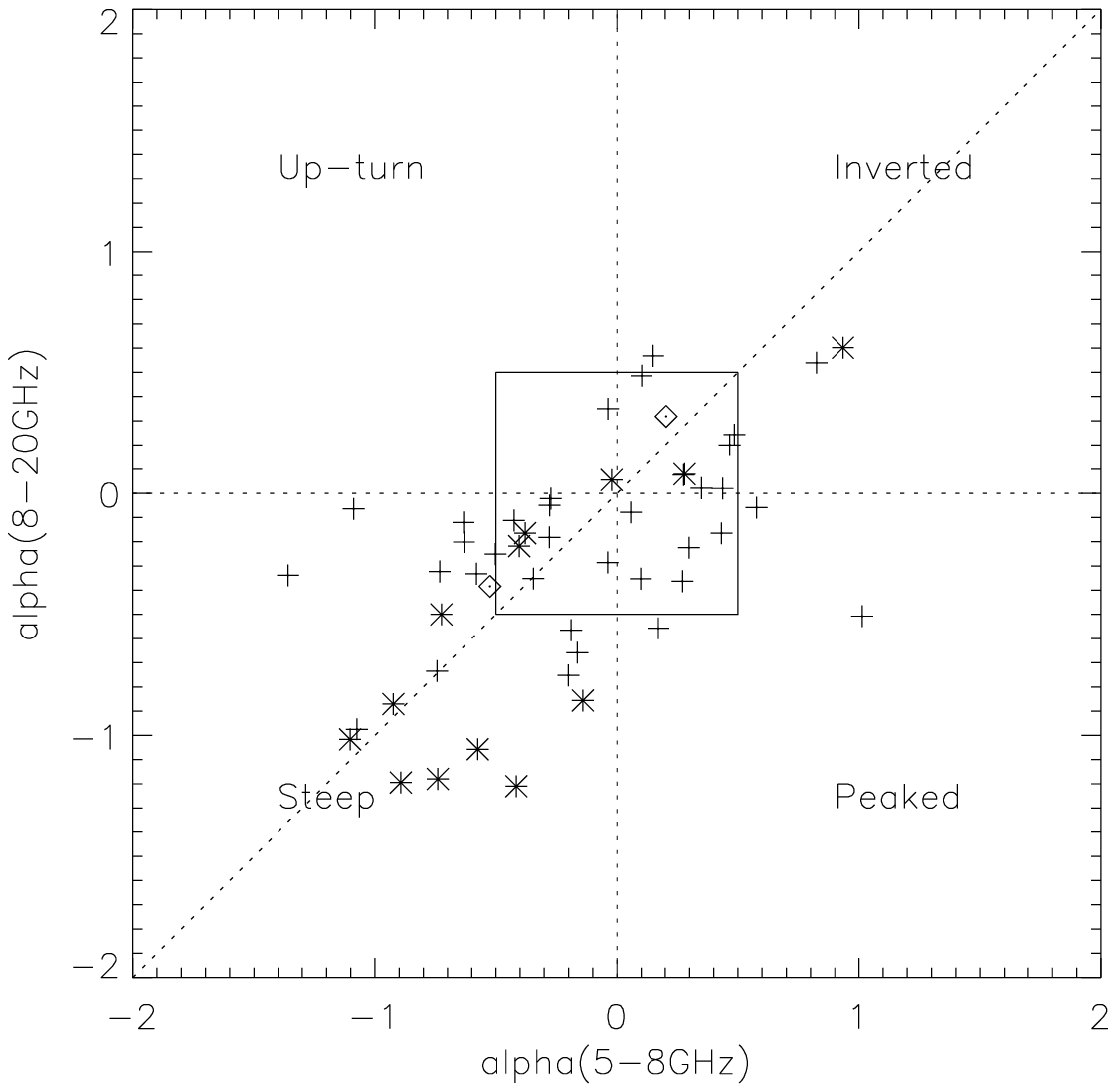}
\end{center}
 \caption{Colour-colour plot ($\alpha_{8}^{20}$ vs $\alpha_{5}^{8}$) with the five-population scheme described in the paper. 
The spectral index error bars have the size of the symbols. The central square confines 
the region of flat-spectrum sources and the diagonal dotted line corresponds to sources 
with single power-law radio spectra. QSOs are indicated by crosses, radiogalaxies by 
stars and sources without classification by diamonds.} \label{fig:3flux}
\end{figure}

In Fig.~\ref{fig:3flux} we show the colour-colour plot divided according to this five-population 
scheme for the 50 sources in our sample with flux density measurements at 5, 8 and 20 GHz. 

In Table~\ref{tab:at20g} we show the comparison between the 
statistics of our sample and the ones in the AT20G FSR Catalog (Massardi et al. 2011a) 
and the PACO Bright sample (Massardi et al. 2011b) for which the same scheme was adopted.
It is apparent that the flat-spectrum and steep-spectrum sources are the 
most represented in both our sample and the AT20G sample with the Full Sample 
(S$_{\rm lim} > 50$ mJy) providing the best match with our statistics. 
The comparison with the PACO Bright sample (Massardi et al. 2011b),
estimated with the source selection applied here to the spectral indices
calculated on their best epoch fit of the SED between 4.5 and 40 GHz,
clearly shows the higher fraction of flat spectrum sources dominating the
population of sources selected above 200 mJy, while at our flux density
limit the steep spectrum objects become more significant. The fractions in
the remaining sub-populations are comparable. It is also worth noticing that 
all the remaining sources in the up-turn quadrant are classified as flat in Table~\ref{tab:at20g}.

\section[]{SUMMARY} \label{sez:sum}

In this paper we presented the results of the multi-frequency follow-up observations
of the K-band Northern Wide Survey (KNoWS) Pilot project. 73 objects were observed 
at 5, 8 and 20 GHz with the 32-m Medicina radio telescope and 30 GHz with the Toru\'n radio 
telescope. 16 objects have a counterpart in the 30 GHz Planck ERCSC. Planck and Toru\'n
30 GHz flux densities are in reasonable agreement, indicating that our flux densities 
are not systematically biased.   

The objects were identified through NED and SuperCosmos as radiogalaxies
(27 per cent) and QSOs (68 per cent) in agreement with previous works (e.g. Sadler et al. 2006).
Two objects have no optical ID and one is a Planetary Nebula.

42 (58 per cent) out of the 72 extragalactic objects have a redshift measurement:
30 are QSOs and 12 are radiogalaxies. For 34 objects (24 QSOs and 10 radiogalaxies)
we computed the optical and radio luminosity. As expected, QSOs are found on 
average at higher redshift than radiogalaxies and are intrinsically brighter 
both in the radio and the optical band.

Four spectral indices in four contiguous frequency ranges were calculated:
$\alpha_{1.4}^{5}$; $\alpha_{5}^{8}$; $\alpha_{8}^{20}$ and $\alpha_{20}^{30}$. 
A clear steepening trend is visible both in the mean and median of the spectral indices when 
moving from lower to higher frequency ranges, a behaviour that is in agreement
with what was found by Massardi et al. (2011a) for the AT20G Full Sample 
Release.

We used radio colour-colour plots to compare the radio spectral properties
of radio sources of different populations with the ones found in other
high-frequency selected catalogues. Radiogalaxies mostly appear in the 
two lower quadrants (corresponding to the steep-spectrum and GPS-like classes)
while QSOs are more scattered around the centre of the diagram indicating 
a prevalent flat radio spectrum: QSOs are therefore more easily singled out
in high frequency selected surveys.


Our sample appears to be in substantial agreement with the AT20G FSR source
population statistics when the latter sample is cut to the same flux density 
limit (100 mJy). Conversely, both our sample and the AT20G FSR statistics 
differ from the one of the Sadler et al. (2006) sample for the up-turn, 
peaked and inverted spectral types even if all the samples have the same 
flux density threshold.

              
\section*{Acknowledgments}

This work is based on observations performed with the Medicina Telescope,
operated by INAF -- Istituto di Radioastronomia. We gratefully thank the 
staff of the Medicina radio telescope for the valuable support provided.
RV used a grant of the ESTRELA (Early-Stage TRaining site for 
European Long-wavelength Astronomy) Network during her PhD. RV thanks 
the staff of the Toru\'n Observatory (Poland) for their kind help with
the 30~GHz follow-up observations and data reduction. 
We gratefully acknowledge the financial support of the Royal Society Paul
Instrument Fund which allowed us to construct the 30 GHz OCRA-p receiver.
We are also grateful to the Polish Ministry of Science and Higher
Education (grant number N N203 390434).
We acknowledge the fundamental support provided by the European Community
through the framework program FP5 (FARADAY) and the contribution of RADIONET
in the technological development of the Italian K-band Multifeed receiver. 
This research has made use of the NASA/IPAC Extragalactic Database (NED) 
which is operated by the Jet Propulsion Laboratory, California Institute of 
Technology, under contract with the National Aeronautics and Space Administration.   
This research has made use of data obtained from the SuperCOSMOS Science Archive, 
prepared and hosted by the Wide Field Astronomy Unit, Institute for Astronomy, 
University of Edinburgh, which is funded by the UK Science and Technology Facilities Council.
The Authors thank the Referee Merja Tornikoski for the useful comments.



\label{lastpage}

\end{document}